\documentstyle[12pt,aasms4]{article}

\def\arcsecpoint{$''\!.$}

\lefthead{Kraemer, Crenshaw, Peterson, \& Filippenko}
\righthead{Narrow-Line Region of NGC 5548}

\begin{document}

\title{Evidence for a Physically Compact Narrow-Line Region
in the Seyfert 1 Galaxy NGC 5548\altaffilmark{1}}

\author{Steven B. Kraemer\altaffilmark{2,3},
D. Michael Crenshaw\altaffilmark{2,4},
Alexei V. Filippenko\altaffilmark{5},\\
and Bradley M. Peterson\altaffilmark{6}}

\altaffiltext{1}{Based on observations with the NASA/ESA {\it Hubble Space 
Telescope}, which is operated by the Association of Universities for Research in 
Astronomy, Inc., under NASA contract NAS5-26555.}

\altaffiltext{2}{Catholic University of America,
NASA/Goddard Space Flight Center, Code 681,
Greenbelt, MD  20771.}

\altaffiltext{3}{Email: stiskraemer@yancey.gsfc.nasa.gov.}

\altaffiltext{4}{Email: crenshaw@buckeye.gsfc.nasa.gov.}

\altaffiltext{5}{Department of Astronomy, University of California, Berkeley,
CA 94720-3411, alex@astro.berkeley.edu.}

\altaffiltext{6}{Department of Astronomy, The Ohio State University,
Columbus, OH 43210, peterson@astronomy.ohio-state.edu.}

\begin{abstract}

We have combined {\it HST}/FOS and ground-based spectra of the Seyfert 1 galaxy 
NGC~5548 to study the narrow emission lines over the 1200 -- 10,000~\AA\ 
region. All of the spectra were obtained when the broad emission line and 
continuum fluxes were at an historic low level, allowing us to accurately 
determine the contribution of the narrow-line region (NLR) to the emission 
lines. We have generated multicomponent photoionization models to investigate 
the relative strength of the high ionization lines compared to those in Seyfert 
2 galaxies, and the weakness of the narrow Mg~II $\lambda$2800 line.

We present evidence for a high ionization component of NLR gas that is very 
close to the nucleus ($\sim$1 pc). This component must be {\it optically thin} 
to ionizing radiation at the Lyman edge (i.e., $\tau_{0}$ $\approx$ 2.5) to 
avoid producing [O~I] and Mg~II in a partially ionized zone. The very high 
ionization lines (N~V, [Ne~V], [Fe~VII], [Fe~X]) are stronger than the 
predictions of our standard model, and we show that this may be due to 
supersolar abundances and/or a ``blue bump'' in the extreme ultraviolet 
(although recent observations do not support the latter). An outer 
component of NLR gas (at only $\sim$70 pc from the continuum source) is needed 
to produce the low ionization lines. We show that the outer component may 
contain dust, which further reduces the Mg II flux by depletion and by 
absorption of the resonance photons after multiple scatterings.

We show that the majority of the emission in the NLR of NGC 5548 must arise 
within about $\sim$70 pc from the nucleus. Thus, the NLR in this Seyfert 1 
galaxy is very physically compact, compared to the typical NLR in Seyfert 2 
galaxies.

\end{abstract}

\keywords{galaxies: individual (NGC 5548) -- galaxies: Seyfert}

\section{Introduction}

NGC 5548 is a bright, low-redshift (z $=$ 0.0172) Seyfert 1 galaxy that has 
received a considerable amount of attention over the past decade. In particular, 
NGC 5548 has been the subject of a number of intensive spectroscopic monitoring 
campaigns in the optical and UV (Korista et al. 1995). 
These efforts have yielded important results on the nature of the continuum 
source and the size, geometry, and kinematics of the broad-line region (BLR), 
for which the responsivity peaks at about 20 light days from the nucleus
(Peterson et al. 1994). Little attention has been paid to the narrow-line region 
(NLR) in NGC 5548, however, although it could lead to a better understanding of 
the circumnuclear environment at much greater distances from the nucleus of this
otherwise well-studied active galaxy.

In general, studies of the NLR in active galaxies are important for 
understanding the nature of the NLR clouds and the interaction of the central 
continuum source with the surrounding galaxy on large scales. Emission-line 
studies and detailed photoionization modeling are particularly useful 
for determining the range of physical conditions and reddening amongst the NLRs 
in active galaxies. A comparison of the NLR properties in Seyfert 
1 and Seyfert 2 galaxies should be helpful in testing unified theories, which 
postulate that the two types are the same object viewed from 
different perspectives, such that the continuum source and BLR are ``hidden'' in 
Seyfert 2 galaxies (Miller \& Goodrich 1990; Antonucci 1993). If this basic 
hypothesis is correct, then the intrinsic properties of the NLRs in Seyfert 1 
and Seyfert 2 galaxies should {\it not} show large systematic differences.

The narrow lines in Seyfert 1 galaxies are more difficult to measure 
than in Seyfert 2 galaxies, due to blending with the broad lines. However, given 
spectra with sufficient signal-to-noise ratio and spectral resolution, these 
components can be isolated and measured reasonably well
(Crenshaw \& Peterson 1986). Measurements of the narrow lines in the optical are 
given by Cohen (1983) for a large number of Seyfert 1 galaxies. In the UV, these 
measurements were difficult with the low spectral resolution of the 
{\it International Ultraviolet Explorer} ({\it IUE}), but are possible with 
instruments on 
the {\it Hubble Space Telescope} ({\it HST}), such as the Faint Object 
Spectrograph (FOS).

The FOS UV spectra of NGC 5548 presented in a previous paper (Crenshaw, Boggess, 
\& Wu 1993; hereafter Paper I) provide a good starting point for detailed 
studies of the NLR in a Seyfert 1 galaxy. These observations happened to occur
at a time when the broad emission lines (and continuum fluxes) were at an 
historic low in the UV, and the contrast between the broad and narrow 
components is thereby enhanced. Paper I gives the UV spectrum and measurements 
of the broad and narrow lines in NGC 5548. Relative to the other narrow lines, 
C~IV $\lambda$1549 is much stronger in NGC 5548 than in Seyfert 2 galaxies, 
indicating a higher ionization parameter and/or harder continuum in the NLR of 
NGC 5548. Narrow Mg II $\lambda$2800 emission is very weak or absent in NGC 
5548, and Paper I presents two possible explanations: 1) the NLR clouds lack the 
presence of a partially-ionized zone (i.e., they are optically thin to ionizing 
radiation), and/or 2) dust grains are present in the NLR clouds, and the Mg II 
flux is weak due to depletion and/or destruction from multiple scatterings and 
eventual absorption of the photons by dust (Kraemer \& Harrington 1986; Ferland 
1992).

We now have the opportunity to investigate the preliminary results from Paper I 
in more detail, by including ground-based optical spectra and photoionization 
models. From the ground-based monitoring campaigns, we have selected 
spectra that cover the full optical range (3000 -- 10,000 \AA) and were
observed around the same time period as the UV data, when the broad-line fluxes 
were very low. The combination of optical and UV lines provides a wide range of 
emission-line diagnostics, as well as an opportunity to deredden the lines using 
the He~II recombination lines. We can then use multicomponent 
photoionization models to match the dereddened line 
ratios and probe the physical conditions in the NLR of NGC 5548.

\section{Observations and Data Analysis}

\subsection{UV and Optical Spectra}

We obtained the FOS UV spectra of NGC 5548 through a 1\arcsecpoint0 circular 
aperture on 1992 July 5 UT. Paper I gives the details of the observations and 
measurements, along with the UV spectrum and emission-line fluxes (with 
associated errors). The observations were made prior to the installation of 
COSTAR on {\it HST}, so near-simultaneous {\it IUE} spectra were used to adjust 
the absolute flux levels of the FOS spectra. As noted in Paper I, 
the scale factors needed to bring the FOS continuum fluxes up to the {\it IUE} 
levels are around 1.4 -- 1.5, which are somewhat higher than the 
values of 1.1 -- 1.3 for our other Seyfert observations. We concluded that the 
Seyfert nucleus may not have been accurately centered in the aperture. Koratkar 
et al. (1996) suggest that another possible explanation for the discrepancy in 
absolute fluxes is nonlinearity in the {\it IUE} detectors. However, we have 
seen no evidence for this possibility in other observations at these flux 
levels, so we have no reason to distrust the {\it IUE} fluxes. In addition, 
reprocessed versions of these spectra that we obtained from the {\it HST} and 
{\it IUE} archives have not changed the original fluxes by more than 10\%, so we 
continue to use the values from Paper I. Some of the emission lines in Paper I 
have only a single number quoted for the flux (as opposed to separate values for 
the broad and narrow components); a single value represents the narrow-line 
contribution, since the broad component is either not present or too weak to be 
detected in these cases.

We selected two optical spectra obtained during a four-year monitoring campaign 
on NGC 5548 (Peterson et al. 1994), from a time interval of $\sim$30 days when 
the H$\beta$ and continuum light curves were at their lowest levels to date.
The spectra were chosen on the basis of their large wavelength coverage
(3000 - 10000 \AA), high signal-to-noise ratio ($\geq$50 per resolution 
element in the continuum at 5200 \AA), and acceptable resolution ($\sim$8 
\AA). The spectra were obtained through a 4\arcsecpoint0 x 
10\arcsecpoint0 aperture with the 3.0-m Shane telescope $+$ Kast spectrograph on 
1992 April 21 and 1992 May 23 UT. Additional details on the observations are 
given by Peterson et al. (1994). The absolute flux levels were 
adjusted by scaling the optical spectra so that the [O~III] $\lambda$5007 flux 
is 5.58 x 10$^{-13}$ ergs s$^{-1}$ cm$^{-2}$, a value determined from 
observations through large apertures on spectrophotometric nights (Peterson et 
al. 1991). The scale factors we used are 1.36 for the 1992 April 24 spectrum and 
1.01 for the 1992 May 23 spectrum.

Plots of the optical spectra are shown in Figure 1 (the UV spectrum is shown in 
Paper I). The contrast between the broad and narrow components of the permitted 
emission lines is most clearly seen in H$\beta$. The 1992 April 21 spectrum was 
obtained at an historic low level, with a continuum flux of  
F$_{\lambda}$(5100~\AA) $=$ 5.5 x 10$^{-15}$ ergs s$^{-1}$ 
cm$^{-2}$ \AA$^{-1}$ and total H$\beta$ flux of F(H$\beta$) $=$ 3.2 x 10$^{-13}$ 
ergs s$^{-1}$ cm$^{-2}$. The flux levels are a little higher for the 1992 May 23 
spectrum with F$_{\lambda}$(5100 \AA) $=$ 6.0 x 10$^{-15}$ ergs s$^{-1}$ 
cm$^{-2}$ \AA$^{-1}$ and F(H$\beta$) $=$ 3.7 x 10$^{-13}$ ergs s$^{-1}$ 
cm$^{-2}$.  At the time of the FOS 
observations on 1992 July 5, the continuum and H$\beta$ fluxes were close to the 
same levels as those from the second optical spectrum, according to the light 
curves of Peterson et al. (1994).

Although the optical aperture is much larger than the one used for the UV 
observations, we have substantial evidence that it does not contain much 
additional NLR flux. Unfortunately, there are no {\it HST} narrow-band images in 
[O~III] or other strong lines that could be used to directly determine the 
distribution of narrow-line emission close to the nucleus. However, there is 
significant evidence that the apparent size of the NLR is very small in NGC 
5548. Peterson et al. (1995) find from a ground-based image that the [O~III] 
emission is pointlike, given a point-spread function that is characterized by a 
width of 2\arcsecpoint0 (FWHM). In addition, Wilson \& Ulvestad (1982) show that 
in an aperture that is 4\arcsecpoint2 in diameter, the [O~III] $\lambda$5007 
fluxes at positions offset from the nucleus by 4\arcsecpoint5 -- 6$''$
are about 100 times weaker than the nuclear flux.
More importantly, in Paper I we found that the strongest UV lines in the {\it 
IUE} 20$''$ x 10$''$ aperture have fluxes that are only $\sim$ 20\% higher than 
those in the FOS 1\arcsecpoint0 aperture. Thus, the observed UV to optical line 
ratios that we quote are at most 20\%  too low, which has little effect on our 
comparisons with the model results.

In order to measure the flux of each narrow optical line, we used
a local baseline determined by linear interpolation between adjacent 
continuum regions or broad profile wings (in the case of profiles 
consisting of broad and narrow components). For severely blended lines like 
H$\alpha$ and [N~II] $\lambda\lambda$6548, 6584, we used the [O~III] 
$\lambda$5007 profile as a template to deblend the lines (see Crenshaw \& 
Peterson 1986). The adopted flux for each narrow component is the average of the 
values from each of the two spectra.

We determined the reddening of the narrow emission lines from the He~II 
$\lambda$1640/$\lambda$4686 ratio and the Galactic reddening curve of Savage \& 
Mathis (1979). For the temperatures and densities typical of the NLR, the He II 
lines are due to recombination, and this particular ratio only varies from 6.3 
to 7.6 (Seaton 1978); we adopt an intrinsic value of 7.2, consistent with our 
model values (Section 3). The observed He~II $\lambda$1640/$\lambda$4686 ratio 
is 5.5 $\pm$ 1.6, which yields a reddening of E$_{B-V}$ $=$ 0.07 mag 
$^{+0.09}_{-0.06}$. The portion of the reddening that is due to our own Galaxy 
is E$_{B-V}$ $=$ 0.03 mag, determined from a neutral hydrogen column density of 
N$_{HI}$ $=$ 1.6 x 10$^{20}$ cm$^{-2}$ (Murphy et al. 1996) and the relationship 
E$_{B-V}$ $=$N$_{HI}$/5.2 x 10$^{21}$ cm$^{-2}$ (Shull \& Van Steenburg 1985).
We note that the intrinsic reddening of the narrow emission lines in this 
Seyfert 1 galaxy, E$_{B-V}$ $\approx$ 0.04 mag, is much smaller than typical 
values of 0.2 -- 0.4 mag obtained for Seyfert 2 galaxies (MacAlpine 1988; 
Ferland \& Osterbrock 1986; Kraemer et al. 1994). 

We determined errors in the dereddened ratios from the sum in quadrature of the 
errors from three sources: photon noise, different reasonable continuum 
placements, and reddening. Errors in the optical ratios are dominated by 
continuum placement, whereas errors in the UV to optical ratios are due to both 
continuum placement and uncertainties in the reddening correction. Errors in the 
weak lines in both regions also have a significant contribution from photon 
noise. As we discussed earlier in this section, there are some possible sources 
of systematic error in the UV to optical line ratios, on which we placed upper 
limits of $\sim$20\%.

Table 1 gives the observed and dereddened narrow-line ratios relative to 
H$\beta$, and errors in the dereddened ratios. Cohen (1983) gives the next 
most comprehensive list of optical line ratios; in general, Cohen's observed 
ratios agree with ours to within the errors. A number of investigators have 
independently determined the narrow H$\beta$/[O~III] $\lambda$5007 ratio in NGC 
5548 (Cohen 1983; Crenshaw \& Peterson 1986; Peterson 1987; Wamsteker et al. 
1990; Rosenblatt et al. 1992; Wanders \& Peterson 1996): these values range from 
0.10 to 0.15, compared to our value of 0.12 $\pm$ 0.01.

\subsection{The Ionizing Continuum}

Estimates of the ionizing continuum are needed as input values for the 
photoionization models of the NLR. We choose the continuum data points given by 
Krolik et al. (1991), since they represent the historic mean levels for this 
object. As always, the greatest uncertainty is the shape of the extreme 
ultraviolet (EUV) continuum. Figure 2 
gives the UV continuum point closest to the EUV region, at 1340 \AA, and the 
X-ray continuum points from Krolik et al. (cf. Turner 
\& Pounds 1989; Clavel et al. 1991) in terms of luminosity (ergs 
s$^{-1}$ Hz$^{-1}$), which we have adjusted for a Hubble constant of H$_{0}$ $=$ 
75 km s$^{-1}$ Mpc$^{-1}$. The dotted line in Figure 1 gives Krolik et al.'s 
continuum fit in the EUV, which is a power law determined from the UV data along 
with an exponential cutoff designed to meet the first X-ray point. We prefer a 
fit with two power laws (in the form L$_{\nu}$ $=$ K$\nu^{\alpha}$), given the 
evidence for an upturn in the spectrum at energies smaller than 1 -- 2 keV 
(i.e., a soft X-ray excess). A fit to these data yields  $\alpha$ $=$ 
$-$1.40$\pm$0.03 in the EUV and soft X-rays, and $\alpha$ $=$ $-$0.40$\pm$0.03 
in the hard X-rays; the break point is at $\nu$ $=$ 10$^{17.1}$ Hz$^{-1}$ (1.3 
keV).

NGC 5548 was monitored by the {\it Extreme Ultraviolet Explorer} ({\it EUVE}) 
over a two month period during 1993 March -- May 
(Marshall et al. 1997).  During this time, the EUV flux varied by a factor of 
four from peak to minimum, and the average flux (corrected for Galactic 
neutral hydrogen absorption) was 135 $\mu$Jy at $\sim$76 \AA.
These observations provide an important constraint on the EUV ionizing 
continuum, but were not used directly in our continuum fit for two reasons. 
First, the neutral hydrogen absorption due to our Galaxy is well known (see the 
previous section), but there could be additional absorption along the 
line of sight. Second, the {\it EUVE} flux, averaged over two months, may not 
be representative of the average flux over many years. Given these caveats, we 
plot the {\it EUVE} continuum point in Figure 2 for comparison with our adopted 
continuum; the error bar was determined from an estimate of $\pm$10\% 
uncertainty in 
Galactic N$_{HI}$ (see Murphy et al. 1996). The {\it EUVE} point is slightly 
higher than 
the continuum fits in Figure 2, but appears to be consistent with our and Krolik 
et al.'s 
adoption of a relatively steep continuum. If we use Krolik et al.'s continuum or 
a continuum formed by joining the UV, {\it EUVE}, and X-ray points with line 
segments 
(in log space), the total EUV flux increases by factors of only 1.19 and 1.30, 
respectively, and the flux at the frequency of the {\it EUVE} observation 
increases by 
factors of 1.40 and 1.84, respectively. The effects of adopting these other 
continua are small, and will be discussed later in the paper.

\section{Photoionization Models}

In modeling the narrow-line emission of NGC 5548, we have adhered to our basic 
philosophy of keeping the number of free parameters to a minimum, by using the 
available observational constraints and the simplest assumptions possible.
The parameters are varied until the best agreement is obtained with the 
observed line ratios, and additional input parameters are only included if 
they are needed to provide a reasonable match to the majority of the  
lines. Discrepancies between the model predictions and specific lines are then 
investigated to provide further insight into the physical conditions. In some 
cases we generate variations on the standard model using additional parameters 
(such as dust) or nonstandard values of the initial parameters (such as nonsolar 
abundances) to illustrate our ideas for resolving the discrepancies.

\subsection{Methodology}

The basic modeling methodology that we employ is described in 
Kraemer et al. (1994) and the details of the photoionization code are given in 
Kraemer (1985). To review the major points, we assume plane parallel 
geometry, which is reasonable if the gas is ionized by radiation from a central 
source at a distance that is large compared to the extent of the cloud. The gas 
is assumed to be atomic (i.e., there is no molecular component). For radiation 
bounded models, we stop the integration into the slab when the electron 
temperature falls below 5000 K and there is no longer any significant line 
emission. The emission line photon escape is through the ionized face of the 
slab. Details of the treatment of dust in the models are described in Kraemer 
(1985). Since the work of Kraemer et al. (1994), we have added iron to the 
elements modeled in the code. The atomic data that we used can be found in 
Pradhan \& Peng (1994), and references contained therein, as well as through 
Ferland's ``Cloudy and Associates'' World Wide Web site 
(http://www.pa.uky.edu/$\sim$gary/cloudy).
The final output of these models is an emission line spectrum. The line
strengths are tabulated relative to H$\beta$. In addition, the model
gives the volume emissivity of H$\beta$, from which we can determine
the mass of gas required to produce the observed line emission, the
efficiency of production of H$\beta$ photons, and an estimate of the covering 
factor.

In order to keep the input parameters to a minimum, we kept two of them fixed 
for our standard model: the shape of the ionizing continuum and the abundances. 
We used the simplest possible ionizing continuum consistent with the 
observations, as described in Section 2.2. In addition, we have assumed 
solar elemental abundances for the standard model as follows (see Lambert \& 
Luck 1978): He = 0.1, C = 3.4 x 10$^{-4}$, O = 6.8 x 10$^{-4}$, N = 1.2 x 
10$^{-4}$, Ne = 1.1 x 10$^{-4}$, S = 1.5 x 10$^{-4}$, Si = 3.1 x 10$^{-5}$, Mg = 
3.3 x 10$^{-5}$, Fe = 4 x 10$^{-5}$ (relative to hydrogen by number).

Our photoionization models are parameterized in terms of the density of
atomic hydrogen (N$_{H}$) and the dimensionless ionization parameter at the 
illuminated face of the cloud:

\begin{equation}
U = \int^{\infty}_{\nu_0} ~\frac{L_\nu}{h\nu}~d\nu ~/~ (4\pi~D^2~N_{H}~c),
\end{equation}

\noindent where {\it L$_{\nu}$} is the frequency-dependent luminosity of 
the ionizing continuum, {\it D} is the distance between the cloud and the 
ionizing source, and h$\nu_{0}$ = 13.6 eV. 

We show that we must add two enhancements to our standard model to obtain an 
acceptable match to the observations. First, we need two components of gas,  
characterized by different ionization parameters and densities. Second, we show 
that the inner component must be optically thin (i.e., radiation bounded) at the 
Lyman edge (13.6 eV). We are then able to vary the ionization parameter and 
density of each component to match the observations. Of course, the resulting 
standard model is an oversimplification, since it is likely that the NLR clouds 
are characterized by a number of different ionization parameters, densities, and 
optical depths. We have effectively averaged the initial conditions for each of 
these components to fit the largest selection of line ratios. Given this 
simplification, the two-component model gives a surprisingly good fit 
to the observations.

The narrow-line spectrum of NGC 5548 is dominated by high ionization lines,
such as C~IV $\lambda$1549, N V $\lambda$1240, and [Ne V] $\lambda\lambda$
3346, 3426, as well as the coronal lines of [Fe~VII] and [Fe~X]. This indicates 
that there is a high ionization component relatively near the central source, 
but presumably outside the BLR, since these lines are much narrower
($\leq$~500~km~s$^{-1}$ FWHM) than the broad lines ($\sim$ 5000 km s$^{-1}$ 
FWHM).
Also present in the spectrum are relatively strong lines of [N II] 
$\lambda\lambda$6584, 6548 and [O II] $\lambda$3727. These are typical of the 
narrow emission line regions of many Seyfert 2 galaxies (Koski 1978; Shuder
\& Osterbrock 1981), and presumably arise in a component of relatively low 
ionization and low density.

The choice of density for these models is based on the relative strengths
of certain forbidden emission lines in the spectrum of NGC 5548. 
An emission line will not be an important coolant in gas with sufficiently 
large electron density that collisional de-excitation of the line will dominate
over radiative transition (see Osterbrock 1989).  For example, the observed 
[Ne~V]~$\lambda$3426/H$\beta$
ratio indicates that the density of the component in which that emission
originates must be less than $\sim$5~x~10$^{7}$ cm$^{-3}$ (DeRobertis \& 
Osterbrock 1984). Given the hardness of the ionizing continuum, one would expect 
high temperatures in this highly ionized gas.
The lower limit on the density of this component can be estimated from the
relative strength of the [O~III] $\lambda$5007 emission. The ratio of [O III] 
$\lambda$5007/ [Ne V] $\lambda$3426
is less than 5, which indicates either low density, highly ionized gas, as
is often seen in the extended NLR of Seyfert 2 galaxies (Storchi-
Bergmann et al. 1996), or density greater than 10$^{6}$ cm$^{-3}$ and 
some collisional supression of the $\lambda$5007 line. The ratio of [O~III] 
$\lambda$4363/$\lambda$5007 is very high ($\sim$0.09). In the low density 
limit, this ratio is
$\sim$7 x 10$^{-3}$ at a temperature of 10$^{4}$~K (Osterbrock 1989). 
At this low density, it is unlikely that 
temperatures consistent with photoionization equilibrium can increase this
ratio to the observed value, and probable that the large 
$\lambda$4363/$\lambda$5007
ratio is due to the fact that some of this emission arises in gas of high
density (i.e., $>$ 10$^{6}$ cm$^{-3}$). 
At this density and level of ionization, the 
[N~II]~ $\lambda\lambda$6548, 6584 emission from this component will be 
negligible, so there must be a lower ionization and lower density component 
present. In order for the [N~II] lines to be among the principal coolants from 
this component, its density must be less than 1 x 10$^{5}$ cm$^{-3}$. Other 
studies (cf. Filippenko \& Halpern 1984; Filippenko 1985; Kraemer et al. 1994) 
have shown that a range of densities in the NLR of Seyfert galaxies is likely, 
so it is not surprising that this condition exists in NGC 5548.

To summarize, the range in density, along with the presence of emission lines 
from a wide range of ionization states, indicates that more than one model 
component is needed to fit the NLR spectrum of NGC 5548. 
The existence of strong high ionization lines such as C IV 
$\lambda$1549, N V $\lambda$1240, and [Ne V] $\lambda\lambda$3346, 3426, and our 
evidence for high densities in the region that they are produced, requires a 
component of gas relatively close to the central source. The weakness of Mg II 
$\lambda$2800 and [O I] $\lambda\lambda$6300, 6364 indicate that these gas 
clouds lack a significant partially ionized zone, and therefore must be
{\it optically thin} to the ionizing radiation (or ``matter bounded''). 
We further investigate the conditions in these two components below.

\subsection {Model Results and Comparison to Observations}

Our approach in modeling NGC 5548 was to fit the high ionization
component first and then add components as needed to fit the lower
ionization lines (in the end, only one additional component was needed). Given 
the constraints and assumptions described in the
previous section, we arrived at values of N$_{H}$ $=$ 1 x 10$^{7}$ cm$^{-3}$ and
U $=$ 10$^{-1.5}$ for the high ionization component. Substantially lower 
densities would result in 
[O~III]~$\lambda$5007 being too strong, and higher densities would quench the 
[Ne V] emission. A higher ionization parameter is possible but, given our EUV 
continuum, would not increase the relative strengths of any of the high 
ionization lines other than [Fe X] $\lambda$6374, at the expense of putting this 
component at distances much closer than $\sim$1 pc from the continuum source 
(see Section 4). Models were run to varying optical depth at the Lyman limit 
$\tau_{0}$, with the constraint that the Mg II and [O I] lines could not become 
too strong. After comparing the results of models run with $\tau_{0}$ $=$ 1.5 to 
10, we found that $\tau_{0}$ $=$ 2.5 gave the best fit. The emission line 
spectrum from this model, INNER, is given in Table 2.

A second component, OUTER, was needed to fit the lower ionization lines. 
We found that U $=$ 10$^{-2.5}$ and N$_{H}$ $=$ 2 x 10$^{4}$ cm$^{-3}$ gave a 
good simultaneous fit to the [O~III]~$\lambda$5007/[O~II]~$\lambda$3727 and 
[N~II] $\lambda$6584/H$\beta$ ratios. Unlike our models for Mrk 3 and I~Zw~92 
(Kraemer \& Harrington 1986; Kraemer et al. 1994), there was no need to add a 
third component for NGC 5548, since there is no obvious contribution from a 
component of very low density ($<$ 10$^{3}$ cm$^{-3}$) low 
ionization gas, such as very strong [O II] $\lambda$3727 and [N~I] $\lambda$5200 
lines. The resulting emission line spectrum from OUTER is also included in 
Table 2. (We will discuss two variations on INNER and OUTER in the next 
section.)

In order to fit the observed (and dereddened) narrow-line spectrum of NGC 5548, 
we combined the output spectrum of the two standard components INNER and OUTER. 
In previous studies, we attempted to weigh the contributions from each component 
to fit specific emission line ratios. For the model of NGC 5548, we simply took 
an equal contribution from INNER and OUTER. The relative simplicity
of the narrow-line spectrum and lack of a strong contribution from very 
low-ionization gas makes such a simple fit possible. The combined spectrum
is given, along with the dereddened observed spectrum for comparison, in
Table 3 (horizontal lines in the table indicate that the models do not predict 
the strengths of these emission lines). 

Comparison of the model predictions to the dereddened observed spectrum in Table 
3 shows agreement, to within the errors, for most of the lines. In particular, 
these include C~IV~$\lambda$1549, He II 
$\lambda$1640, [O~II] $\lambda$3727, [Ne III] $\lambda$3869, [N II] 
$\lambda$6584, and the Balmer decrement.
In radiation bounded gas,
the ratio of the He II lines to H$\beta$ is strongly dependent on the shape of 
the ionizing continuum, because neutral hydrogen is the 
dominant absorber of ionizing radiation between 13.6 eV and 54.4 eV, while above 
54.4 eV, singly ionized helium dominates.
If there is a component of matter bounded gas, He II/H$\beta$ is less easily 
predicted. The accuracy of our fit to this ratio indicates that the relative 
contributions of the matter and radiation bounded components are approximately 
correct, given the observational constraints on the ionizing continuum.
The fact that we have a reasonable fit 
for lines that span a wide range of ionization and critical densities supports 
our values for density and ionization parameter. Most of the 
discrepancies between the observations and models are in the lowest and highest 
ionization lines, which we will address below. 

\subsection{Discrepancies and Possible Explanations}

First, we address differences between the predicted and observed ratios for the 
low ionization lines. The Mg II $\lambda$2800 and [O I] $\lambda\lambda$6300, 
6364 lines are still predicted to be too strong by our standard model, by 
factors $\geq$ 6 and 2.5, respectively. Nearly all this emission is coming from 
OUTER. Two factors determine the strength of the [O I] lines: the hardness of 
the ionizing continuum and the physical depth of the emission line clouds. Since 
it appears that we have a good fit for the ionizing continuum, the weakness in 
the observed [O~I] lines gives a limit on the depth of the clouds. Truncating 
the integration of OUTER at $\tau_{0}$ $\approx$ 1000 would give a better fit to 
the [O I] without affecting the other important line ratios. This results in a 
cloud depth of $\sim$2.5 x 10$^{15}$ cm.
The overprediction of the Mg II $\lambda$2800 line strength presents a somewhat
different problem. In order to reduce the contribution of this line from INNER, 
we assumed a matter bounded model for this component. To provide a better match 
to the observation of little or no Mg II emission, we can reduce the model 
contribution by modifying OUTER. The Mg$^{+}$ emissivity is greatest near the 
H$^{+}$/H$^{0}$ transition zone in OUTER, so a 
simple truncation at much lower optical depths is not feasible, as it would have
a much greater effect on the other line ratios. 

An obvious explanation for the weak observed Mg II is depletion of the magnesium 
into dust grains, along with suppression of the resonance photons by multiple 
scatterings and eventual absorption by dust. This was suggested in Paper I (cf. 
Kraemer \& Harrington 1986; Ferland 1992). For comparison with our standard 
model, we generated a version of OUTER that includes dust, assuming 
a dust to gas ratio that is 30\% of that found in the Galactic interstellar 
medium, with equal amounts of graphite and silicate grains and accompanying 
depletions. These assumptions were made to avoid biasing our results by simply 
having all of the Mg depleted into dust grains. We assumed relative element 
depletions as calculated by Seab \& 
Shull (1983), and the grain size distributions determined by Mathis et al. 
(1977) and Draine \& Lee (1984); details of the treatment of dust in the code 
are given by Kraemer (1985). The results of the model are given in Table 2, and 
not only show a significant drop in the relative strength of Mg II 
$\lambda$2800, but also a drop in the Ly$\alpha$ strength, as is expected due to 
the preferential dust absorption of multiply scattered UV resonance lines. The 
lower Ly$\alpha$/H$\beta$ ratio is a better fit to the observations. A 
substantially larger dust-to-gas ratio than we assumed would result in a 
Ly$\alpha$/H$\beta$ ratio that is lower than observed. Therefore, 
it is likely that there is some dust mixed in with the low-ionization gas, 
although with a lower dust to gas ratio than found in the ISM, and that 
depletion coupled with the resonance line suppression explain the weak Mg II. 

Second, we address discrepancies in the high ionization lines.
Specifically, the lines of N~V, [Ne~V], [Fe~VII], and [Fe~X] are too weak by 
factors of 2 to 4 compared to the observations.
As we mentioned earlier, the density is well constrained, so increasing the 
ionization parameter brings the gas well within 1 pc, into the realm of 
the BLR. However, these lines are relatively narrow (FWHM $\leq$ 500 km 
s$^{-1}$, see Moore et al. 1996) and they are not likely to arise very close to 
the BLR. The strengths of the high ionization lines can also be enhanced 
relative to H$\beta$ by truncating the integration of INNER at a lower optical 
depth. However, this has the problem of enhancing the He II emission relative to 
H$\beta$ in the model. More likely solutions to the problem of underpredicting 
the high ionization lines include 1) shock ionization, 2) a large ``blue 
bump'' in the EUV continuum, or 3) supersolar abundances.

Predicting the strengths of the coronal lines has always been a problem with 
simple photoionization models, as Viegas-Aldrovandi \& Contini (1989) discuss in 
some detail for the Fe lines. They suggest that there may be shocked gas mixed 
in with the photoionized clouds and that these high ionization lines may arise 
there. Although this is certainly a possible factor, there may be other 
plausible explanations which avoid adding another level of complexity.

An obvious way in which the coronal lines might be enhanced is if there
were a component of ionizing radiation that contributed significantly
at energies between 100 and 500 eV. Although there has been some
speculation about the presence of a ``blue bump'' in the EUV, recent work
by Zheng et al. (1997) on low-redshift quasars shows that the {\it near} EUV 
continuum is likely to be much steeper than previously 
supposed. In NGC 5548, the {\it EUVE} continuum point in Figure 2 is further 
confirmation that a large blue bump is not present in the spectrum 
of NGC 5548. Another possibility may be diffuse radiation from the intercloud 
medium. Tran (1995) has shown that there may be a contribution to the continuum 
radiation in some Seyfert 2 galaxies from thermal emission from the intercloud
medium responsible for the scattering of the hidden BLR emission into
the observer's line of sight. If a similar medium with temperatures
$\approx$ 5~x~10$^{5}$~K exists in the NLR of NGC 5548, it is possible that 
free-free radiation and line emission that arise within it may contribute to the 
ionization of the inner narrow-line gas. Although this component would be weak 
compared to the continuum radiation emitted by the central source, it could have
a significant local contribution to the ionization balance of clouds existing
within this inner region. However, recent observations suggest that the extended 
UV continuum seen in some Seyferts may be due to starbursts (Heckman et al. 
1997), which would not contribute to the high ionization lines.

In studies of medium redshift QSOs, Ferland et al. (1996) found
evidence of supersolar abundances. There is no direct evidence of
elemental enhancements in Seyfert galaxies, but it is certainly not
implausible, particularly near the nucleus, where the most intense activity
occurs. As Oliva (1996) points out, the coronal line emission will be enhanced 
proportionally to the abundance of the atomic species.
As a comparison, we ran a version of INNER with a heavy element
abundance that is twice solar, and the results are shown in Table 3. In 
particular,
the relative strengths of the [Fe VII] and [Ne V] lines have increased,
while many lines, such as C IV $\lambda$1549 and C III] $\lambda$1909, show 
little change. It is possible, then, that the observed strength of 
some of these lines is in part due to enhanced abundances. Note that we have not 
attempted to adjust the increase in abundances to fit assumptions about the
type of star formation that might be expected.

\section{Discussion}

From our standard model, we can estimate several global properties of the NLR in 
NGC 5548, including the covering factor and physical size, in addition to more 
local properties, including optical depth and presence of dust.
We determine the covering factor of the NLR gas from the observed and model 
values of the ``conversion efficiency'', $\eta$ , which is the ratio of H$\beta$ 
photons to ionizing photons. The covering factor is given by
C~$=$~$\eta$(observed)/$\eta$(model). A value of C $>$ 1 would indicate that the 
ionizing radiation is anisotropic, which we would not expect to be the case 
for a Seyfert 1 galaxy, since the central source is seen directly in such 
objects. Assuming Ho= 75 km s$^{-1}$ Mpc$^{-1}$, the observed H$\beta$ 
flux corresponds to a luminosity of 3.8 x 10$^{40}$ ergs s$^{-1}$, or 9.3 x 
10$^{51}$  H$\beta$ photons s$^{-1}$. From the continuum observations described 
in section 2.2, we calculate a total luminosity of ionizing photons of 1.09 x 
10$^{54}$ s$^{-1}$. This yields an observed $\eta$ = 0.009. Our fit to the 
observed emission line spectrum assumed that each component in our model 
contributed 50\% of the H$\beta$ emission. The resulting values of $\eta$ were 
0.06 for INNER and 0.11 for OUTER, and the covering factors are 0.07 and 0.04, 
respectively, so C(NLR) = 0.11. The value is small compared to those found for 
Seyfert 2 galaxies, which are often $>$ 1 (Kinney et al. 1991; Kraemer et al. 
1994), and there is no evidence for anisotropic radiation. Note that this 
estimate of covering factor does not include the BLR clouds, which contribute at 
least 50\% of the flux in many of the strong lines, even when NGC 5548 is in its 
lowest state (Paper I). 

Given the ionization parameters and densities of the two components from our 
standard model, as well as the ionizing luminosity, the 
characteristic sizes (i.e., radii) for the two emitting regions are 1 pc for 
INNER, and 70 pc for OUTER. 
(Using the higher continuum luminosity given by the {\it EUVE} point in Figure 2 
would increase these values by a factor of only $\sqrt{1.30}$, or 1.14.)
Thus, the NLR of NGC 5548 is {\it physically} 
compact, since the size of OUTER is much smaller than typical values of 200 -- 
1000 pc determined for Seyfert 2 galaxies, using the same methods that we have 
described in this paper (Kraemer \& Harrington 1986; Kraemer et al. 1994).
Although there have been reports of extended emission from NGC 5548 (Wilson et 
al. 1989), the contribution to the integrated emission line spectrum from this 
region is small (Peterson et al. 1995); we estimated the contribution to the 
narrow UV lines outside of a 1$''$ aperture (330 pc for H$_{0}$ $=$ 75 km 
s$^{-1}$ Mpc$^{-1}$) to be only $\sim$20\%. This is consistent with our finding 
that the majority of the narrow emission must arise in a region with a 
``diameter'' of 140 pc.

Pogge (1989) and Schmitt \& Kinney (1996) claim that the
{\it apparent} size of the NLR in a Seyfert 1 galaxy is typically much 
smaller than that of a Seyfert 2 galaxy, although this may be due to a selection 
effect, since most of the Seyfert 2s in the {it HST} archive were selected on 
the basis of their extended emission (Wilson 1997).
Nevertheless, it is clear that the majority of the Seyfert 1 galaxies in these 
studies are apparently compact. This cannot be explained 
by viewing angle alone, since the opening angles 
of the presumed ionization cones are large, and in the simplest version of the 
unified model, the apparent extent of most Seyfert galaxies should be much 
larger, even if viewed ``pole-on'' (Schmitt \& Kinney 1996).
Our results show that a possible explanation for the small apparent size of the 
NLRs in many Seyfert 1 galaxies is that they are truly (i.e., physically) 
compact. Schmitt \& Kinney explain this phenomenon as a result of
the orientation of the obscuring torus with respect
to the plane of the galaxy. However, their model does not explain the dominance
of the high ionization lines in NGC 5548 and many other Seyfert 1 galaxies, a 
feature not generally seen in Seyfert 2 galaxies 
(Koski 1978; Shuder \& Osterbrock 1981). Thus, if the unified model applies we 
might expect that the narrow emission line spectrum of this Seyfert 1 galaxy 
would resemble that of Seyfert 2 galaxies, and, further, we would expect 
to see a noticeable contribution to the spectrum from gas hundreds of parsecs 
from the central source. It is possible that the high ionization region in 
Seyfert 2 galaxies is obscured by a torus, but this does not explain the absence 
of a low ionization component in NGC 5548. Not only do our models show that the 
narrow-line spectrum can be well fitted without such a component, but it is 
clear that the NLR of NGC 5548 is dominated by high ionization gas that must 
be located close to the central source.

A few Seyfert 1 galaxies do indeed show significant NLR emission at
large distances from the central source. For example, {\it HST} observations of 
NGC 4151 (Evans et al. 1993; Hutchings et al. 1997) reveal an array of knots and 
filaments out to a kiloparsec that are almost certainly ionized by the central 
continuum. It may be that some Seyfert 1 galaxies, such as NGC 4151, have 
extended emission line regions and narrow lines in their spectra that closely
resemble Seyfert 2 galaxies. Cohen (1983) studied the optical narrow
line spectra of a group of Seyfert 1 galaxies and found some resemblance
although, as a group, they appeared to be of somewhat higher ionization.
Certainly, some of the galaxies in his study have spectra that are
indistinguishable from those of type 2 Seyferts, but others, like NGC 5548, 
appear to be dominated by high ionization lines, a condition that appears to be 
rare among Seyfert 2 galaxies. It would be extremely interesting if one could 
determine if these high ionization objects are as compact as NGC 5548 appears to 
be. Unfortunately, optical spectra alone are insufficient for detailed 
modeling of the NLR, and thus far, accurate measurements of the narrow-line 
strengths in the UV have only been obtained for NGC 5548
and NGC 4151 (Ferland \& Mushotzky 1982).

As we stated above, our finding that the inner component of gas is
optically thin at the Lyman limit is based both on the absence of strong
Mg II emission and the relative strength of the high ionization lines.
If this is indeed true, not only for NGC 5548 but for other Seyfert 1
galaxies, it may be a clue to the origin of the inner narrow-line
gas. Thin filaments or knots of reasonably high density ($\sim$ 10$^{7}$ 
cm$^{-3}$) could be the result of outflow from the BLR, either as condensations
in an expanding intercloud medium or as ``tails'' of BLR clouds, driven
out by radiation pressure. If so, the small physical depths of the clouds
inferred from our modeling ($\sim$ 10$^{14}$ cm) may constrain the sizes of 
their BLR progenitors. Note that there has been some recent success in 
models of the BLR using a component of
optically thin clouds (Shields, Ferland, \& Peterson 1995).
There have been other studies suggesting that the 
clouds in the NLR are matter bounded or that some mix of radiation bounded and 
matter bounded clouds exist (Wilson et al. 1997; Viegas-Aldrovandi 1988). Such a 
mix might give rise to the filamentary structure seen in some [O III] images of 
Seyfert galaxies rather than the typical molecular clouds found in spiral 
galaxies, and may give a clue about the origin of the NLR gas.

One other point regarding the NLR gas is that it appears that there
may be some dust present within the outer components of the emitting
region. Dust is certainly able to exist in clouds at this proximity
to the central source, with dust temperatures reaching a few hundred
K (Kraemer \& Harrington 1986). Nevertheless, the history of the dust
in this gas is unknown. Our finding that the outer clouds
may not be purely radiation bounded would indicate that they are not
simply interstellar molecular clouds that have an outer shell ionized
by the central source. Our finding that there is probably dust mixed in with
this gas will be important in determining the origin of this component.

Finally, Moore et al. (1996) find a correlation of ionization potential of the 
narrow emission lines with velocity width in NGC 5548. Combined with our finding 
that, to first order, the ionization level decreases with distance, this 
indicates that the radial velocities decrease with increasing distance. This 
trend is also seen in a more direct fashion in spatially resolved spectra of the 
inner NLR in NGC 4151 (Hutchings et al. 1997).

\section{Conclusions}

We have analyzed UV and optical spectra of the Seyfert 1 galaxy NGC 5548 that 
were obtained when it was at an historical minimum. We were able to 
isolate the narrow emission line components (which do not vary in flux on short 
time scales), due to the relative weakness of the more rapidly variable broad 
emission lines that are usually blended with the narrow components.
We have constructed photoionization models of the narrow-line region of this 
galaxy, and are able to successfully match the observed dereddened ratios of a 
large number of emission lines to within the errors, with the 
exceptions noted in Section 3.3. Since we used the best direct observational 
evidence of the shape of the ionizing continuum, rather than making adjustments 
based on fitting emission line ratios, the quality of this fit is particularly 
satisfying. The fact that a good fit was obtained for the permitted emission 
lines, such as C~IV~$\lambda$1549, and the forbidden lines, such as [Ne III] 
$\lambda$3869, [O III] $\lambda$5007, [O II] $\lambda$3727, [N II] 
$\lambda$6584, etc., indicates that the range of physical conditions assumed 
in these models is approximately correct.

From our analysis and modeling of these spectra, we can make several statements 
regarding the physical conditions in the NLR of NGC 5548. First, it is clear 
that the principal source of ionization in the NLR of NGC 5548 is the central 
continuum source. This conclusion is borne out by the quality of the fit to the 
emission line spectrum. The NLR covering factor is reasonably small (C $=$ 
0.11), so the NLR gas does not need to intercept much of the ionizing continuum 
to produce the observed emission lines fluxes. A second conclusion is that the 
highly ionized gas in the inner part of the NLR appears to be optically thin at 
the Lyman limit ($\tau_{0}$ $\approx$ 2.5), which yields constraints on the 
physical depths of these clouds and may provide a clue to their origin. We have 
also presented evidence for supersolar abundances in the inner portion of the 
NLR, and dust in the outer portion. 

The most important conclusion that we have reached in this study is
that the NLR of NGC 5548 is physically compact, with the majority of emission 
coming from a distance $\leq$~70 pc from the nucleus. By contrast, the unified 
model of Seyfert galaxies suggests that the {\it physical} dimensions of the NLR 
in Seyfert 1 and 2 galaxies should be similar. Additional studies of the type 
that we have presented in this paper, particularly of other Seyfert 1 galaxies, 
are important for testing this aspect of the unified model.

\acknowledgments

We thank the referee, Andrew Wilson, for helpful comments and suggestions.
We are grateful to Fred Bruhweiler and Pat Harrington for helpful discussions on 
the physical properties of iron and the availability of atomic data.
S.B.K. and D.M.C. acknowledge support from NASA grant NAG~5-4103.
A.V.F acknowledges support from NASA grant NAG~5-3556, and B.M.P. 
acknowledges support from NSF grant AST-9420080.

\clearpage

\figcaption[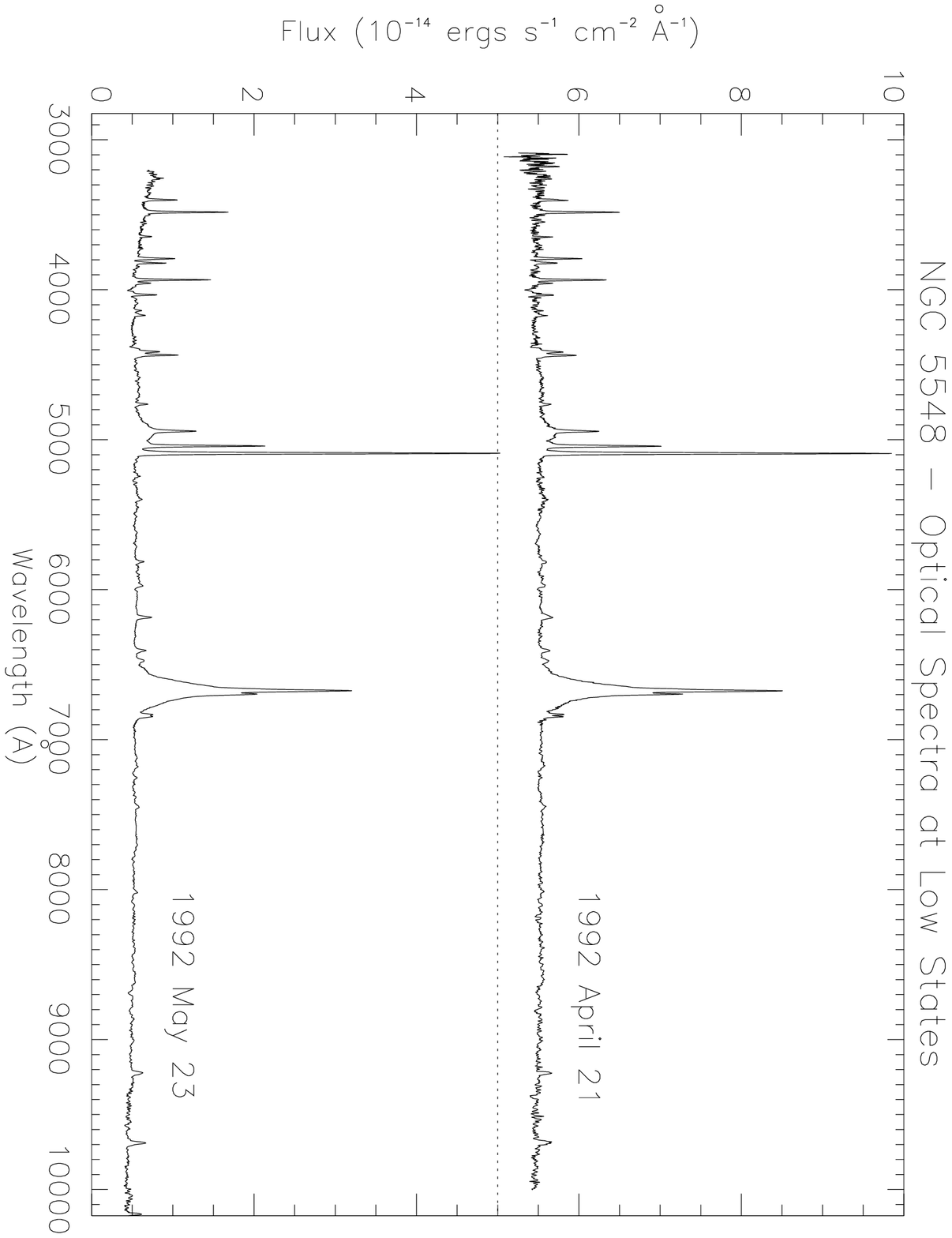]{Optical ground-based spectra of NGC 5548, obtained
when the continuum and broad-line fluxes were at a very low level.
The upper spectrum is offset by a constant flux of 5 x 10$^{-14}$ ergs s$^{-1}$ 
cm$^{-2}$ \AA$^{-1}$.
}\label{fig1}

\figcaption[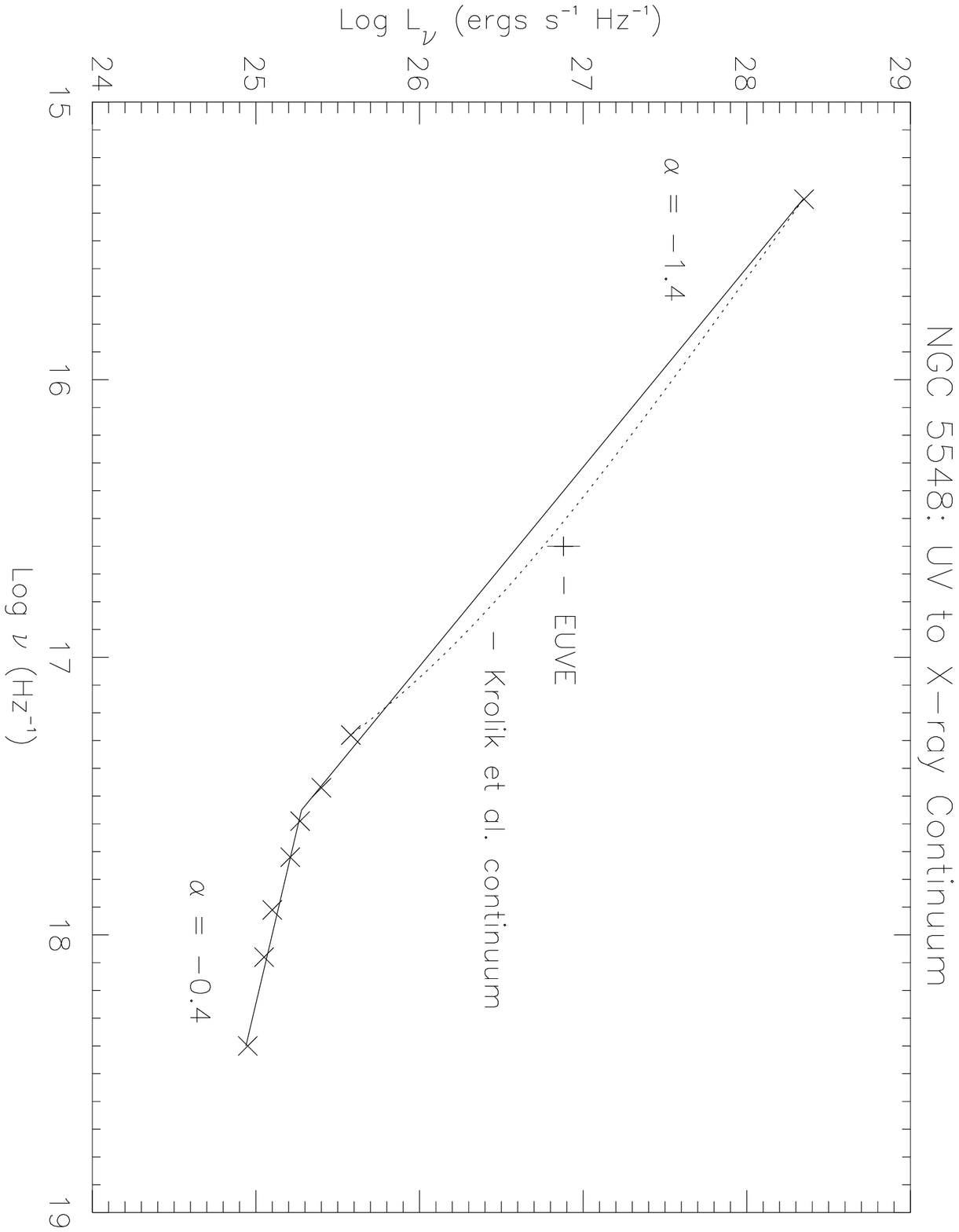]{UV to X-ray continuum of NGC 5548, in luminosity 
assuming H$_{0}$ $=$ 75 km s$^{-1}$ Mpc$^{-1}$. Observed UV and X-ray points and 
dotted line fit are from Krolik et al. (1991). Our broken power-law fit 
($\alpha$ $=$ $-$1.4, $-$0.4) is given by the solid line. The {\it EUVE} 
continuum point described in the text is given for comparison.
}\label{fig2} 
 
\clearpage
\begin{deluxetable}{lrrr}
\tablecolumns{4}
\footnotesize
\tablecaption{Narrow-line ratios for NGC 5548 
(relative to H$\beta$$^{a}$)\label{tbl-1}}
\tablewidth{0pt}
\tablehead{
\colhead{} & \colhead{} & \colhead{Reddening$^{b}$} & \colhead{}\\
\colhead{} & \colhead{Observed} & \colhead{Corrected} & 
\colhead{~~~~~~~~~~~~$\sigma$ $^{c}$} 
}
\startdata
Ly$\alpha$ $\lambda$1216     &14.60  &22.01  &($\pm$7.53) \\ 
N V $\lambda$1240            &1.82   &2.70   &($\pm$0.95) \\
O I $\lambda$1302            &0.81   &1.15   &($\pm$0.60) \\
Si IV/O IV] $\lambda$1400   &2.10   &2.88   &($\pm$0.99) \\
N IV] $\lambda$1486          &0.30   &0.40   &($\pm$0.15) \\
C IV $\lambda$1550           &10.73  &14.22  &($\pm$4.10) \\ 
He II $\lambda$1640          &1.22   &1.60   &($\pm$0.50) \\
O III] $\lambda$1663         &0.63   &0.82   &($\pm$0.26) \\
Si III] $\lambda$1892        &0.24   &0.32   &($\pm$0.13) \\
C III] $\lambda$1909         &2.18   &2.91   &($\pm$0.83) \\
O III]/C II]$\lambda$2323    &0.37   &0.51   &($\pm$0.23) \\
$[$Ne IV] $\lambda$2423      &0.24   &0.31   &($\pm$0.15) \\
$[$O II] $\lambda$2470       &0.21   &0.27   &($\pm$0.11) \\
Mg II $\lambda$2800          &$<$0.15 &$<$0.18 &     \\
O III $\lambda$3133          &0.45   &0.50   &($\pm$0.18) \\
He II $\lambda$3204          &0.15   &0.17   &($\pm$0.07) \\
$[$Ne V] $\lambda$3346       &0.57   &0.62   &($\pm$0.13) \\ 
$[$Ne V] $\lambda$3426       &1.70   &1.84   &($\pm$0.30) \\
$[$Fe VII] $\lambda$3588     &0.21   &0.23   &($\pm$0.07) \\
$[$O II] $\lambda$3727       &0.79   &0.84   &($\pm$0.14) \\  
$[$Fe VII] $\lambda$3760     &0.45   &0.48   &($\pm$0.08) \\ 
$[$Ne III] $\lambda$3869     &1.25   &1.32   &($\pm$0.18) \\ 
H$\zeta$ + He I $\lambda$3889 &0.27  &0.28   &($\pm$0.18) \\
$[$Ne III] + H$\epsilon$ $\lambda$3967 &0.42   &0.44   &($\pm$0.07) \\   
\tablebreak
$[$S II] $\lambda$4072       &0.12   &0.13   &($\pm$0.03) \\ 
H$\delta$ $\lambda$4102      &0.21   &0.22   &($\pm$0.05) \\ 
H$\gamma$ $\lambda$4340      &0.45   &0.46   &($\pm$0.09) \\
$[$O III] $\lambda$4363      &0.75   &0.77   &($\pm$0.11) \\
He II $\lambda$4686          &0.22   &0.23   &($\pm$0.05) \\ 
H$\beta$ $\lambda$4861       &1.00   &1.00   & \\  
$[$O III] $\lambda$4959      &2.61   &2.60   &($\pm$0.36) \\ 
$[$O III] $\lambda$5007      &8.13   &8.07   &($\pm$0.88) \\ 
$[$Fe VII] $\lambda$5721     &0.21   &0.20   &($\pm$0.05) \\  
He I $\lambda$5876           &0.22   &0.21   &($\pm$0.05) \\
$[$Fe VII] $\lambda$6087     &0.48   &0.45   &($\pm$0.08) \\
$[$O I] $\lambda$6300        &0.33   &0.31   &($\pm$0.07) \\
$[$O I] $\lambda$6364        &0.11   &0.10   &($\pm$0.03) \\
$[$Fe X] $\lambda$6374       &0.18   &0.17   &($\pm$0.06) \\
$[$N II] $\lambda$6548       &0.27   &0.25   &($\pm$0.06) \\
H$\alpha$ $\lambda$6563      &3.30   &3.06   &($\pm$0.44) \\  
$[$N II] $\lambda$6583       &0.82   &0.76   &($\pm$0.15) \\
$[$S II] $\lambda$6716       &0.36   &0.33   &($\pm$0.07) \\
$[$S II] $\lambda$6730       &0.36   &0.33   &($\pm$0.07) \\  
$[$O II] $\lambda$7325       &0.21   &0.19   &($\pm$0.06) \\
$[$S III] $\lambda$9069      &0.45   &0.39   &($\pm$0.09) \\
$[$S III] $\lambda$9532      &0.79   &0.68   &($\pm$0.13) \\
\tablenotetext{a}{Flux (H$\beta$) = 6.7 ($\pm$0.7) x 
10$^{-14}$ ergs s$^{-1}$ cm$^{-2}$.}
\tablenotetext{b}{Calculated using E$_{B-V}$ = 0.07 mag.}
\tablenotetext{c}{Estimated uncertainty in the reddening-corrected ratio.}
\enddata
\end{deluxetable}

\clearpage
\begin{deluxetable}{lrrrr}
\tablecolumns{5}
\footnotesize
\tablecaption{Line Ratios from model components\label{tbl-2}}
\tablewidth{0pt}
\tablehead{
\colhead{} & \colhead{INNER$^{a}$} & \colhead{INNER$^{a}$} &
\colhead{OUTER$^{b}$} & \colhead{OUTER$^{b}$} \\
\colhead{} & \colhead{(solar)} & \colhead{(2x solar)} &
\colhead{(no dust)} & \colhead{(dust)}
}
\startdata
C III $\lambda$977                    & 1.16   & 0.70   & 0.03  & 0.03 \\
N III $\lambda$990                    & 0.16   & 0.11	& 0.00  & 0.00 \\
O VI $\lambda$1036                    & 1.43   & 1.12	& 0.00  & 0.00 \\
Si III $\lambda$1206                  & 0.05   & 0.05	& 0.02  & 0.02 \\
O V $\lambda$1216                     & 1.94   & 1.64	& 0.00  & 0.00 \\
Ly$\alpha$ $\lambda$1216    	      &36.24   &35.59	&38.02  &20.26 \\
N V $\lambda$1240           	      & 2.02   & 1.86	& 0.00  & 0.00 \\
C II $\lambda$1334           	      & 0.05   & 0.12	& 0.13  & 0.06 \\
Si IV $\lambda$1398  	              & 0.52   & 0.54	& 0.08  & 0.06 \\
O IV] $\lambda$1402                   & 2.67   & 2.52	& 0.04  & 0.05 \\
S IV] $\lambda$1417                   & 0.14   & 0.13	& 0.01  & 0.01 \\
N IV] $\lambda$1486         	      & 1.91   & 1.69	& 0.03  & 0.03 \\
C IV $\lambda$1550          	      &30.87   &30.08	& 0.66  & 0.45 \\
$[$Ne V] $\lambda$1575                & 0.04   & 0.05	& 0.00 & 0.00 \\
$[$Ne IV] $\lambda$1602               & 0.31   & 0.31	& 0.01  & 0.02 \\
He II $\lambda$1640         	      & 2.64   & 2.39	& 1.57  & 1.71 \\
O III] $\lambda$1663        	      & 2.86   & 2.73	& 0.26  & 0.30 \\
N III] $\lambda$1750                  & 0.92   & 0.99	& 0.16  & 0.19 \\
$[$Ne III] $\lambda$1815              & 0.03   & 0.03	& 0.00  & 0.00 \\
Si III] $\lambda$1883                 & 0.00   & 0.00	& 0.05  & 0.05 \\
Si III] $\lambda$1892       	      & 0.31   & 0.42	& 0.20  & 0.20 \\
C III] $\lambda$1909        	      & 8.08   & 8.78	& 2.19  & 1.98 \\
$[$O III] $\lambda$2321               & 0.51   & 0.59   & 0.04  & 0.05 \\
C II] $\lambda$2326                   & 0.07   & 0.10	& 1.37  & 1.09 \\
\tablebreak
$[$Ne IV] $\lambda$2423     	      & 0.03   & 0.03	& 0.26  & 0.29 \\
$[$O II] $\lambda$2470      	      & 0.01   & 0.02	& 0.29  & 0.31 \\
$[$Mg V] $\lambda$2784                & 0.14   & 0.19	& 0.02  & 0.01 \\
Mg II $\lambda$2800         	      & 0.11   & 0.18	& 2.12  & 0.81 \\
$[$Mg V] $\lambda$2929         	      & 0.04   & 0.05	& 0.00  & 0.00 \\
$[$Ne V] $\lambda$2974                & 0.02   & 0.02	& 0.00  & 0.00 \\
He II $\lambda$3204                   & 0.15   & 0.14	& 0.10  & 0.11 \\
$[$Ne III] $\lambda$3342              & 0.02   & 0.02	& 0.00  & 0.00 \\
$[$Ne V] $\lambda$3346      	      & 0.29   & 0.41	& 0.04  & 0.04 \\
$[$Ne V] $\lambda$3426      	      & 0.79   & 1.11	& 0.10  & 0.10 \\
$[$N I] $\lambda$3467                 & 0.00   & 0.00	& 0.01  & 0.01 \\
$[$Fe VII] $\lambda$3588    	      & 0.11   & 0.16	& 0.01  & 0.00 \\
$[$S III] $\lambda$3722               & 0.01   & 0.01	& 0.04  & 0.04 \\
$[$O II] $\lambda$3727      	      & 0.00   & 0.00	& 1.61  & 1.70 \\
$[$Fe VII] $\lambda$3760    	      & 0.15   & 0.22	& 0.01  & 0.01 \\
$[$S III] $\lambda$3796               & 0.00   & 0.00	& 0.00  & 0.00 \\
$[$Ne III] $\lambda$3869    	      & 1.11   & 1.67	& 1.28  & 1.42 \\
$[$Ne III] $\lambda$3967              & 0.35   & 0.51	& 0.39  & 0.44 \\
$[$S II] $\lambda$4072                & 0.00   & 0.00	& 0.16  & 0.17 \\
H$\delta$ $\lambda$4100 	      & 0.26   & 0.26	& 0.26  & 0.26 \\
H$\gamma$ $\lambda$4340      	      & 0.47   & 0.47	& 0.47  & 0.47 \\
$[$O III] $\lambda$4363      	      & 2.24   & 2.57	& 0.18  & 0.20 \\
He I $\lambda$4471      	      & 0.03   & 0.03	& 0.04  & 0.04 \\
Mg I] $\lambda$4571                   & 0.00   & 0.00	& 0.03  & 0.03 \\
He II $\lambda$4686          	      & 0.36   & 0.33	& 0.23  & 0.25 \\
\tablebreak
$[$Ne IV] $\lambda$4720       	      & 0.07   & 0.08	& 0.00  & 0.00 \\
H$\beta$                 	      & 1.00   & 1.00	& 1.00  & 1.00 \\
$[$O III] $\lambda$5007      	      & 3.02   & 4.43	&17.63  &18.05 \\
$[$N I] $\lambda$5198                 & 0.00   & 0.00	& 0.29  & 0.30 \\
$[$N I] $\lambda$5200                 & 0.00   & 0.00	& 0.23  & 0.23 \\
He II $\lambda$5412                  & 0.03   & 0.03	& 0.02  & 0.02 \\
$[$O I] $\lambda$5577                 & 0.00   & 0.00	& 0.02  & 0.02 \\
$[$Fe VII] $\lambda$5721     	      & 0.15   & 0.23	& 0.02  & 0.02 \\
$[$N II] $\lambda$5755                & 0.00   & 0.00   & 0.05  & 0.05 \\
He I $\lambda$5876           	      & 0.08   & 0.09	& 0.11  & 0.11 \\
$[$Fe VII] $\lambda$6087     	      & 0.22   & 0.34	& 0.03  & 0.03 \\
$[$O I] $\lambda$6300        	      & 0.00   & 0.00	& 1.51  & 1.51 \\
$[$S III] $\lambda$6312               & 0.01   & 0.02   & 0.06  & 0.07 \\
$[$O I] $\lambda$6364        	      & 0.00   & 0.00	& 0.50  & 0.50 \\
$[$Fe X] $\lambda$6374       	      & 0.13   & 0.13	& 0.00  & 0.00 \\
$[$N II] $\lambda$6548       	      & 0.00   & 0.00	& 0.62  & 0.68 \\
H$\alpha$ $\lambda$6563      	      & 2.80   & 2.80	& 2.97  & 3.00 \\
$[$N II] $\lambda$6584       	      & 0.00   & 0.00	& 1.81  & 2.01 \\
$[$S II] $\lambda\lambda$6716, 6731   & 0.00   & 0.00	& 1.19  & 1.18 \\
$[$O II] $\lambda$7325       	      & 0.01   & 0.02	& 0.38  & 0.43 \\
$[$S III] $\lambda$9069      	      & 0.00   & 0.01	& 1.21  & 1.33 \\
$[$S III] $\lambda$9532      	      & 0.01   & 0.02	& 2.95  & 3.24 \\
$[$S II] $\lambda$10,300              & 0.00   & 0.00	& 0.11  & 0.11 \\
$[$N I] $\lambda$10,395               & 0.00   & 0.00	& 0.04  & 0.05 \\
$[$N I] $\lambda$10,404               & 0.00   & 0.00	& 0.03  & 0.03 \\
\tablenotetext{a}{U $=$ 10$^{-1.5}$, N$_{H}$ $=$  1 x 10$^{7}$ cm$^{-3}$,
$\tau_{0}$ = 2.5.}
\tablenotetext{b}{U $=$ 10$^{-2.5}$, N$_{H}$ $=$  2 x 10$^{4}$ cm$^{-3}$.}
\enddata
\end{deluxetable}

\clearpage
\begin{deluxetable}{lrr}
\tablecolumns{3}
\footnotesize
\tablecaption{Line ratios from standard model$^{a}$ and 
observations\label{tbl-3}}
\tablewidth{0pt}
\tablehead{
\colhead{} & \colhead{Model} & \colhead{Dereddened}
}
\startdata
Ly$\alpha$ $\lambda$1216    	       &37.10   &~~~22.01   ($\pm$7.53) \\
N V $\lambda$1240           	       & 1.00   &2.70	 ($\pm$0.95) \\
O I $\lambda$1302           	       & ---    &1.15	 ($\pm$0.60) \\
Si IV/O IV] $\lambda$1400  	       & 1.65   &2.88	 ($\pm$0.99) \\
N IV] $\lambda$1486         	       & 0.97   &0.40	 ($\pm$0.15) \\
C IV $\lambda$1550          	       &15.80   &14.22   ($\pm$4.10) \\
He II $\lambda$1640         	       & 2.11   &1.60	 ($\pm$0.50) \\
O III] $\lambda$1663        	       & 1.56   &0.82	 ($\pm$0.26) \\
Si III] $\lambda$1892       	       & 0.26   &0.32	 ($\pm$0.13) \\
C III] $\lambda$1909        	       & 5.14   &2.91	 ($\pm$0.83) \\
O III]/C II]$\lambda$2323   	       & 1.00   &0.51	 ($\pm$0.23) \\
$[$Ne IV] $\lambda$2423     	       & 0.13   &0.31	 ($\pm$0.15) \\
$[$O II] $\lambda$2470      	       & 0.15   &0.27	 ($\pm$0.11) \\
Mg II $\lambda$2800         	       & 1.12   &$<$0.18	     \\
O III $\lambda$3133         	       & ---    &0.50	 ($\pm$0.18) \\
He II $\lambda$3204         	       & 0.12   &0.17	 ($\pm$0.07) \\
$[$Ne V] $\lambda$3346      	       & 0.16   &0.62	 ($\pm$0.13) \\
$[$Ne V] $\lambda$3426      	       & 0.44   &1.84	 ($\pm$0.30) \\
$[$Fe VII] $\lambda$3588    	       & 0.06   &0.23	 ($\pm$0.07) \\
$[$O II] $\lambda$3727      	       & 0.81   &0.84	 ($\pm$0.14) \\
$[$Fe VII] $\lambda$3760    	       & 0.08   &0.48	 ($\pm$0.08) \\
$[$Ne III] $\lambda$3869    	       & 1.20   &1.32	 ($\pm$0.18) \\
H$\zeta$ + He I $\lambda$3889          & ---    &0.28	 ($\pm$0.18) \\
$[$Ne III] + H$\epsilon$ $\lambda$3967 & 0.37   &0.44	 ($\pm$0.07) \\
\tablebreak
$[$S II] $\lambda$4072       	       & 0.08   &~~~~0.13 ($\pm$0.03) \\
H$\delta$ $\lambda$4102      	       & 0.26   &0.22	 ($\pm$0.05) \\
H$\gamma$ $\lambda$4340      	       & 0.47   &0.46	 ($\pm$0.09) \\
$[$O III] $\lambda$4363      	       & 1.21   &0.77	 ($\pm$0.11) \\
He II $\lambda$4686          	       & 0.29   &0.23	 ($\pm$0.05) \\
H$\beta$ $\lambda$4861       	       & 1.00   &1.00		     \\
$[$O III] $\lambda$4959      	       & 3.44   &2.60	 ($\pm$0.36) \\
$[$O III] $\lambda$5007      	       &10.32   &8.07	 ($\pm$0.88) \\
$[$Fe VII] $\lambda$5721     	       & 0.08   &0.20	 ($\pm$0.05) \\
He I $\lambda$5876           	       & 0.09   &0.21	 ($\pm$0.05) \\
$[$Fe VII] $\lambda$6087     	       & 0.13   &0.45	 ($\pm$0.08) \\
$[$O I] $\lambda$6300        	       & 0.76   &0.31	 ($\pm$0.07) \\
$[$O I] $\lambda$6364        	       & 0.25   &0.10	 ($\pm$0.03) \\
$[$Fe X] $\lambda$6374       	       & 0.07   &0.17	 ($\pm$0.06) \\
$[$N II] $\lambda$6548       	       & 0.31   &0.25	 ($\pm$0.06) \\
H$\alpha$ $\lambda$6563      	       & 2.89   &3.06	 ($\pm$0.44) \\
$[$N II] $\lambda$6583       	       & 0.91   &0.76	 ($\pm$0.15) \\
$[$S II] $\lambda\lambda$6716, 6730    & 0.59   &0.66	 ($\pm$0.10) \\
$[$O II] $\lambda$7325       	       & 0.20   &0.19	 ($\pm$0.06) \\
$[$S III] $\lambda$9069      	       & 0.61   &0.39	 ($\pm$0.09) \\
$[$S III] $\lambda$9532      	       & 1.47   &0.68	 ($\pm$0.13) \\
\tablenotetext{a}{50\% H$\beta$ contribution from INNER (solar abundances) and 
50\% from OUTER (no dust).} 
\enddata
\end{deluxetable}
				       				       
\clearpage
\plotone{fig1.eps}

\plotone{fig2.eps}

\end{document}